\documentclass{ws-ijmpa}
\usepackage{graphicx}
\usepackage[super,compress]{cite}
\usepackage{hyperref}
\hypersetup{colorlinks,urlcolor=black,citecolor=black,linkcolor=black,filecolor=black}
\usepackage{breakurl}

\newcommand{\comment}[1]{}

\newcommand{\eV}{\ensuremath{\text{eV}}}
\newcommand{\meV}{\ensuremath{\operatorname{meV}}}
\newcommand{\lr}[1]{ \left( #1 \right) }
\newcommand{\lrs}[1]{ \left[ #1 \right] }

\newcommand{\vev}[1]{ \langle \, #1 \, \rangle }
\newcommand{\Tr}{ {\rm Tr} \, }
\newcommand{\tr}{ {\rm Tr} \, }

\renewcommand{\det}[1]{ {\rm det} \left( #1 \right) }

\newcommand{\expa}[1]{ \exp{\left( #1 \right)} }

\begin{document}
\markboth{P.~V.~Buividovich, M.~V.~Ulybyshev}{Lattice QCD techniques for condensed matter}

%
\catchline{}{}{}{}{}
%

\title{Applications of lattice QCD techniques for condensed matter systems}

\author{P.~V.~Buividovich}

\address{Institute of Theoretical Physics, University of Regensburg, Universitatsstrasse 31, Regensburg,
D-93053 Germany\\
Pavel.Buividovich@physik.uni-regensburg.de}

\author{M.~V.~Ulybyshev}

\address{Institute of Theoretical Physics, University of Regensburg, Universitatsstrasse 31, Regensburg,
D-93053 Germany\\
Institute for Theoretical Problems of Microphysics, Moscow State University, Moscow, 119899 Russia\\
Maksim.Ulybyshev@physik.uni-regensburg.de}

\maketitle


\begin{abstract}
 We review the application of lattice QCD techniques, most notably the Hybrid Monte-Carlo (HMC) simulations, to first-principle study of tight-binding models of crystalline solids with strong inter-electron interactions. After providing a basic introduction into the HMC algorithm as applied to condensed matter systems, we review HMC simulations of graphene, which in the recent years have helped to understand the semi-metal behavior of clean suspended graphene at the quantitative level. We also briefly summarize other novel physical results obtained in these simulations. Then we comment on the applicability of Hybrid Monte-Carlo to topological insulators and Dirac and Weyl semi-metals and highlight some of the relevant open physical problems. Finally, we also touch upon the lattice strong-coupling expansion technique as applied to condensed matter systems.
\end{abstract}

\ccode{PACS numbers:73.22.Pr, 11.15.Ha}

\section{Introduction}
\label{sec:intro}

 Crystalline materials with low-energy electronic excitations which can be described as Dirac fermions are nowadays in the focus of active theoretical and experimental research. Probably the best and most known example of such a material is provided by graphene, a two-dimensional layer of carbon atoms. More recent examples include Dirac semimetals such as $Cd_3 \, As_2$ \cite{Borisenko:14:1} and $Zr \, Te_5$ \cite{Kharzeev:14:1}, topological insulators such as $B_{1-x} \, Sb_x$ and $Bi_2 \, Se_3$ \cite{Hsieh:09:1} and Weyl semimetals such as $Ta \, As$ \cite{Huang:15:1} \cite{Hasan:15:1}.

 In table-top experiments with such materials one can realize a wealth of transport phenomena and spontaneous symmetry breaking patterns which are (or have been) commonly attributed to the realm of high-energy and nuclear physics. Notable examples include Klein tunnelling \cite{Katsnelson:06:2}, supercritical charge \cite{Katsnelson:07:1}, spontaneous chiral symmetry breaking \cite{Lahde:09:1} and the Chiral Magnetic Effect \cite{Kharzeev:14:1}.

 Advances in experimental studies of these new materials naturally called for advanced first-principle calculation methods, most notably numerical. Provided the parameters of the tight-binding model of the material under consideration are known, say, from Density Functional (DFT) calculations, determinantal Quantum Monte-Carlo simulations are suitable for obtaining fully non-perturbative results with controllable statistical errors.

 Before the graphene era, the central point of application of determinantal Monte-Carlo simulations with fermions was the Hubbard model in $(2+1)$ dimensions, which is believed to describe some high-$T_c$ superconductors \cite{Jarrell:09:05, Varney:09:08}. Since Hubbard model is only a qualitative, not a quantitative, model of high-$T_c$ superconductivity, in numerical studies thereof one is mostly interested in qualitative features of the phase diagram. In contrast to the case of high-$T_c$ superconductors, the tight-binding model of graphene is very simple and is characterized by only a few parameters with well-known values, such that it can be easily implemented in determinantal Monte-Carlo simulations practically without any model assumptions. Furthermore, since graphene is a truly 2D material, Coulomb interactions are only weakly screened. As a result, the long-range tail of the Coulomb potential is physically very important, and interactions can not be described by a single on-site potential - at least the interactions between the electrons on neighboring lattice sites should be considered \cite{Schuler:13:1}. On the one hand, the presence of strong long-range interactions makes even the mean-field calculations rather complicated, not to mention the Diagrammatic Monte-Carlo methods \footnote{see however \cite{Sazonov:15:1} for some recent attempts}. On the other hand, it precludes the use of the discrete Hubbard-Stratonovich transformation \cite{Hirsch:83:1}, which is very often a method of choice for models with on-site interactions. Yet another difficulty is that long-range interactions typically require very large system size in order to fully control the finite-volume effects, which rules out the methods based on the direct evaluation of fermionic determinants or ratios thereof.

 For all these reasons, it seems that the Hybrid Monte-Carlo (HMC) algorithm \cite{MontvayMuenster,DeGrandDeTarLQCD}, commonly used in lattice QCD simulations, is a very convenient method for studying the physics of Dirac materials, most notably graphene. First, it has no difficulty addressing the effect of interactions with arbitrarily long range. Second, it does not require an explicit calculation of fermionic determinant, which allows one to study very large systems at very low temperatures (e.g. up to $48 \times 48$ lattice cells for graphene, with temperature as low as $0.1 \eV$). Finally, at zero chemical potential it is free of the sign problem for most tight-binding models of Dirac quasiparticles with particle-hole symmetry. It is interesting that an early attempt to apply HMC to condensed matter systems was made already in the eighties \cite{Scalettar:86:1}, but the method only showed its full potential when applied to graphene much later.

 In this paper, we will review the most important aspects, results and limitations of the Hybrid Monte-Carlo simulations of graphene, and consider the possible applications of HMC to other Dirac materials. In Section \ref{sec:hmc_intro} we start with a brief introduction into the HMC algorithm which can be used for simulations of condensed matter systems. We continue with a brief review of the HMC simulations of graphene along with the physical problems which motivated them, from the first attempts with staggered fermions to the most recent simulations with screened Coulomb potential. After that, in Section \ref{sec:applications} we provide an overview of the most interesting results in graphene physics obtained so far using HMC simulations. In Section \ref{sec:outlook} we conclude with an outlook for the application of HMC and also lattice strong coupling expansion technique to other Dirac materials.

\section{Hybrid Monte-Carlo algorithm for condensed matter systems}
\label{sec:hmc_intro}

 The starting point for the formulation of the HMC algorithm is the interacting Hamiltonian with particles and holes as fermionic degrees of freedom:
\begin{eqnarray}
\label{fermionic_hamiltonian1}
 \hat{H} =
 \sum\limits_{x, y} \hat{\psi}^{\dag}_x h^{\psi}_{xy} \hat{\psi}_y
 +
 \sum\limits_{x, y} \hat{\chi}^{\dag}_x h^{\chi}_{xy} \hat{\chi}_y
 +
 V_{x y} \hat{q}_x \hat{q}_y .
\end{eqnarray}
Here $\hat{\psi}_x^{\dag}$, $\hat{\psi}_x$ and $\hat{\chi}_x^{\dag}$, $\hat{\chi}_x$ are the creation and annihilation operators for particles and holes, respectively, obeying the usual fermionic anti-commutation relations, $x$ and $y$ label lattice sites and $h^{\psi, \chi}_{x y}$ are the single-particle Hamiltonian operators for particles and holes. $V_{x y}$ is the arbitrary long-range interaction potential between electric charges on lattice sites $x$ and $y$, which are defined as $\hat{q}_x = \hat{\psi}^{\dag}_x \hat{\psi}_x - \hat{\chi}^{\dag}_x \hat{\chi}_x$. After the standard Suzuki-Trotter and Hubbard-Stratonovich transformations, one can represent the partition function $\mathcal{Z}  = \tr \expa{-\hat{H}/T}$ in terms of the path integral over the real-valued, single-component Hubbard-Stratonovich field $\phi_x\lr{\tau}$:
\begin{eqnarray}
\label{partition_function_HS}
 \mathcal{Z} = \int \mathcal{D}\phi_x\lr{\tau} \,
 \det{M^{\psi}} \, \det{M^{\chi}} \,
 \expa{-\frac{1}{2} \, \int\limits_{0}^{T^{-1}} d\tau \sum\limits_{x,y} \phi_x\lr{\tau} V^{-1}_{x y} \phi_y\lr{\tau}  } ,
\end{eqnarray}
where $\tau$ is the ``Euclidean time'' variable taking values in the range $\tau \in \lrs{0, T^{-1}}$, $V^{-1}_{x y}$ is the matrix inverse of the potential $V_{x y}$ defined via the identity $\sum\limits_z V_{x z} V^{-1}_{z y} = \delta_{x y}$ and $M^{\psi, \chi}$ are the fermionic operators which act on time-dependent single-particle wave functions $\Psi_x\lr{\tau}$ as
\begin{eqnarray}
\label{fermionic_matrix_op}
 \lrs{M^{\Psi} \, \Psi}_x\lr{\tau} = \partial_{\tau} \Psi_x\lr{\tau} - \sum\limits_y h^{\Psi}_{x y} \Psi_y\lr{\tau} - i \sigma \phi_x\lr{\tau} \Psi_x\lr{\tau} ,
\end{eqnarray}
where the sign $\sigma = +1$ for particles ($\Psi = \psi$) and $\sigma = -1$ for holes ($\Psi = \chi$).

Different practical discretizations of the Euclidean time $\tau$ were studied in detail in \cite{Smekal:13:1,Smekal:13:3}. Upon discretization, the Hubbard-Stratonovich field $\phi_x\lr{\tau}$ typically enters the fermionic operator as a time-like component of the compact gauge field $\partial_{\tau} \psi_x\lr{\tau} - i \phi_x\lr{\tau} \psi_x\lr{\tau} \rightarrow \lr{e^{i \phi_x\lr{\tau}} \psi_x\lr{\tau + \Delta \tau} - \psi_x\lr{\tau}}/{\Delta \tau}$. Thus in contrast to lattice QCD simulations, in the simulations of condensed matter one typically needs only the time-like component of the gauge field. An important consequence is that the time-reversal symmetry, if present, is unbroken for every configuration of the Hubbard-Stratonovich field.

If the single-particle Hamiltonians $h^{\psi}$ and $h^{\chi}$ are equal to each other (particle-hole symmetry) or have complex conjugate elements $h^{\psi}_{x y} = \bar{h}^{\chi}_{x y}$ then $\det{M^{\psi} M^{\chi}} = \det{M M^{\dag}}$, where $M$ is either $M^{\psi}$ or $M^{\chi}$. Thus the path integral weight in (\ref{partition_function_HS}) is positive and can be used for Monte-Carlo sampling of the Hubbard-Stratonovich field $\phi_x\lr{\tau}$.

 In order to avoid the calculation of the full fermionic determinant in (\ref{partition_function_HS}), in the so-called $\Phi$ HMC algorithm \cite{DeGrandDeTarLQCD} one rewrites the determinant as a Gaussian integral over a pseudo-fermion field $\Phi_x\lr{\tau} \in \mathbb{C}$. For the molecular dynamics step of the HMC algorithm, one also adds the ``momentum'' fields $\pi_x\lr{\tau}$ which are completely decoupled at the level of the path integral, so that the partition function reads
\begin{eqnarray}
\label{partition_function_HMC}
 \mathcal{Z} = \int
  \mathcal{D}\phi_x\lr{\tau}
  \mathcal{D}\pi_x\lr{\tau}
  \expa{ - \frac{\Delta \tau}{2} \, \lr{ \sum\limits_{\tau,x,y} \phi_x\lr{\tau} V^{-1}_{x y} \phi_y\lr{\tau} + \sum\limits_{\tau,x} \pi_x^2\lr{\tau}
  }}
  \times \nonumber \\ \times
  \int
  \mathcal{D}\bar{\Phi}_x\lr{\tau}
  \mathcal{D}\Phi_x\lr{\tau} \,
  \expa{-\sum\limits_{x,y,\tau, \tau'}
  \bar{\Phi}_x\lr{\tau} \lrs{M M^{\dag}}^{-1}_{x y}\lr{\tau \tau'} \Phi_y\lr{\tau'}
  }  .
\end{eqnarray}

 The configurations of the fields $\phi_x\lr{\tau}$ and $\Phi_x\lr{\tau}$ can be now sampled by iteratively repeating the following set of Metropolis transitions:
\begin{enumerate}
 \item \textbf{Random update of the $\Phi$ field:} \, At fixed $\phi$ and $\pi$ fields, generate the new values of the $\Phi$ field with the probability proportional to the exponent in the second line of (\ref{partition_function_HMC}): $\Phi_x\lr{\tau} = \sum\limits_{\tau', y} M_{x y}\lr{\tau, \tau'} \eta_y\lr{\tau'}$, where $\eta_x\lr{\tau} \in \mathbb{C}$ is Gaussian random vector with independent components.
 \item \textbf{Molecular dynamics for the $\phi$ and $\pi$ fields:} \, Generate new values for the momentum field $\pi_x\lr{\tau}$ with the probability $\sim \expa{-\frac{\Delta \tau}{2}\sum\limits_{x,\tau} \pi_x^2\lr{\tau}}$. Then, at fixed pseudo-fermion field $\Phi$, evolve the fields $\phi$ and $\pi$ from $t=0$ up to some finite value of $t \sim 1$ according to the classical equations of motion, which can be obtained by interpreting $\phi_x\lr{\tau}$ and $\pi_x\lr{\tau}$ as canonically conjugate variables and the sum of the arguments of all the exponents in (\ref{partition_function_HMC}) - as a classical Hamiltonian $\mathcal{H}_{cl}\lr{\pi, \phi}$ (with a minus sign). This evolution is parameterized by some fictitious time $t$, and the Euclidean time $\tau$ is treated on equal footing with the spatial coordinates:
     \begin{eqnarray}
     \label{molecular_dynamics}
      \partial_t \phi_x\lr{\tau} = {\Delta \tau} \, \pi_x\lr{\tau},
      \quad
      \partial_t \pi_x\lr{\tau} = - {\Delta \tau} \, \sum\limits_y V^{-1}_{x y} \phi_y\lr{\tau}
      + \nonumber \\ +
\sum\limits_{z, y, \tau', \tau''}\bar{\Psi}_z\lr{\tau'} \frac{\partial \lrs{M M^{\dag}}_{z y}\lr{\tau', \tau''} }{\partial \phi_x\lr{\tau}} \Psi_y\lr{\tau''},
     \end{eqnarray}
     where $\Psi_x\lr{\tau}$ is the solution of the linear system $ \sum\limits_{\tau', y} \lrs{M M^{\dag}}^{-1}_{x y}\lr{\tau, \tau'} \Psi_{y}\lr{\tau'} = \Phi_x\lr{\tau}$. This system should be solved at each step of molecular dynamics evolution, which is typically accomplished by a conjugate gradient method. 
 \item \textbf{Acceptance:} \, In order to correct for discretization errors in the numerical solution of (\ref{molecular_dynamics}) one finally accepts new configuration with the probability $\alpha = \min\lr{1, e^{-\delta \mathcal{H}_{cl}}}$, where $\delta \mathcal{H}_{cl}$ is the change of classical Hamiltonian through the molecular dynamics trajectory. In this case, a discretization of (\ref{molecular_dynamics}) is only required to preserve volume element of the $\lr{\pi, \phi}$ space, which holds for leap-frog \cite{DeGrandDeTarLQCD}, Sexton-Weingarten \cite{Sexton:92:1} or Omelyan integrators \cite{Omelyan:03:1}.
\end{enumerate}

One of the important limitations of the above described algorithm is the case when the matrix of inter-electron interaction potentials $V_{x y}$ is not positive definite, and the Hubbard-Stratonovich representation (\ref{partition_function_HS}) does not exist. Such situation can arise, for example, in mechanically strained graphene \cite{Assaad:15:1}.

\section{Hybrid Monte-Carlo simulations of graphene: motivation and recent developments}
\label{sec:graphene_history}

 Due to the essentially non-relativistic behavior of electrons, Dirac quasiparticles\footnote{It has to be remembered that these Dirac quasiparticles provide only an effective description of the collective motion of electrons in the background of crystalline lattice, and are only very indirectly related to the Dirac Hamiltonian describing the electrons themselves.} in crystalline solids propagate with a Fermi velocity $v_F \sim 10^{-2} c$, rather than with the full speed of light $c$. For this reason, electromagnetic interactions between them can be well approximated simply by instantaneous electrostatic interaction. For the same reason, however, the effective value $\alpha_{eff}$ of the QED coupling constant $\alpha_{QED} \approx 1/137$ which enters the perturbation theory for these quasiparticles is enhanced as $\alpha_{eff} = \alpha_{QED} c/v_F$. E.g. for clean suspended graphene $v_F \approx 1/300$ and $\alpha_{eff} = 2.19$. Such a large value of the effective coupling can potentially lead to numerous non-perturbative phenomena, most notably to spontaneous breaking of the effective chiral symmetry of Dirac quasiparticles, which should result in the opening of the mass gap in the quasiparticle spectrum and hence in the insulating behavior of graphene.

Combined with the extremely high charge carrier mobility \cite{Bolotin:08:2}, the existence of the gap in the quasiparticle spectrum of graphene would enable the development of extremely efficient graphene-based transistors \cite{Nevius:15:1}. Thus it is not surprising that the scenario of spontaneous chiral symmetry breaking in graphene attracted a lot of attention and has been studied using a variety of quantum field theory methods such as Schwinger-Dyson and gap equations \cite{Leal:04:1,Khveshchenko:09:1,Gamayun:10:1,Smekal:13:2,Popovici:13:1}, large-$N$ and strong-coupling expansions \cite{Son:07:1,Drut:08:1} and renormalization group \cite{Vafek:08:1,Gonzalez:10:1} techniques. Different methods predicted quite different values of the critical coupling $\alpha_c$ separating the semimetal and the insulator phases, with most of them, however, lying below $\alpha_{eff} = 2.2$ (see \cite{Wang:12:1} for a nice summarizing table of $\alpha_c$). Thus it was strongly suspected that suspended graphene should be in the insulating phase. Around 2010, experiments could not reliably rule out this possibility (see e.g. \cite{Andrei:08:1,Drut:10:2}).

 This controversy on the insulating or semimetal behaviour of clean graphene was one of the main motivations for first-principle numerical simulations using Monte-Carlo techniques. Since initially it was believed that the insulator-semimetal transition should be most sensitive to the infrared behavior in the vicinity of the Dirac cones, and the full dispersion relation on the hexagonal lattice is of little relevance, the seminal simulations were performed around 2008 using staggered fermions on the square $\lr{2 + 1}$ dimensional lattice coupled to non-compact gauge fields in $\lr{3 + 1}$ dimensions \cite{Lahde:09:1,Lahde:09:2,Lahde:09:3}. Similar simulations with only on-site interactions were performed in \cite{Hands:08:1}. Very fortunately, $N_f = 1$ flavor of $(2+1)$-dimensional staggered fermions corresponds to two physical fermions, which is exactly the number of Dirac cones in graphene. This has allowed to use the lattice QCD simulation codes with staggered fermions with minimal modifications. The outcome of these extensive simulations was the predicted value $\alpha_c = 1.1$ - which implied that clean suspended graphene should be deep in the insulating phase!

 More than two year after the seminal paper \cite{Lahde:09:1}, Manchester group has performed very precise experiments \cite{Elias:11:1,Elias:12:1} which put a very small upper limit $\Delta E \sim 1 \meV$ on the energy gap in clean suspended graphene. It was then immediately clear that simulations with staggered fermions miss some essential physics. At the same time, it was realized that the tight-binding model of graphene on the hexagonal lattice is as well suited for Hybrid Monte-Carlo simulations as $\lr{2 + 1}$ dimensional staggered fermions, but has a clear advantage of having all the correct symmetries \cite{Rebbi:11:1,Rebbi:12:1}. Upon neglecting the next-to-nearest neighbor hoppings, the single-particle Hamiltonians for both particles and holes in graphene can be written in the following simple form:
\begin{eqnarray}
\label{graphene_hamiltonian}
 h_{x y} = -\kappa \sum\limits_{a=1}^3 \delta_{x + \hat{a}, y} ,
\end{eqnarray}
where $\kappa \approx 2.7 \eV$ is the hopping amplitude, $x$ and $y$ label the sites of the hexagonal lattice and $\hat{a}$, $a = 1 \ldots 3$ denote the three unit vectors pointing from the lattice site $x$ to the three closest neighbor sites. The full interacting Hamiltonian takes the general form (\ref{fermionic_hamiltonian1}).

 At the level of the tight-binding model (\ref{graphene_hamiltonian}), the continuum $U\lr{1}$ chiral symmetry of the Dirac Hamiltonian is replaced by the $Z_2$ symmetry which exchanges the components of single-particle wave functions between the two simple triangular sublattices (usually referred to as $A$ and $B$) of the hexagonal lattice. The order parameter which corresponds to the chiral condensate of Dirac fermions can be written as the difference of particles and/or hole numbers between the sublattices:
\begin{eqnarray}
\label{chiral_condensate_graphene}
 \Delta_{\psi}
 =
 \sum\limits_{x \in A} \vev{ \hat{\psi}^{\dag}_x \hat{\psi}_x }
 -
 \sum\limits_{x \in B} \vev{ \hat{\psi}^{\dag}_x \hat{\psi}_x },
\end{eqnarray}
and similarly for holes with creation/annihilation operators $\hat{\chi}_x^{\dag}$ and $\hat{\chi}_x$.

The $U\lr{1}$ chiral symmetry of the single-particle Dirac Hamiltonian is only an approximate, \emph{emergent} symmetry at low energies. This implies that while the $Z_2$ sublattice symmetry can still be spontaneously broken due to sufficiently strong interactions, this breaking is not accompanied by the emergence of the gapless Goldstone modes. For the full Hamiltonian (\ref{fermionic_hamiltonian1}) there is also a continuum $SU\lr{2}$ symmetry which rotates between particles and holes. Different combinations of the particle and hole occupation numbers on the two sublattices correspond to different patterns of the spontaneous breaking of the full $Z_2 \otimes SU\lr{2}$ symmetry of the Hamiltonian (\ref{fermionic_hamiltonian1}) with the single-particle Hamiltonian (\ref{graphene_hamiltonian}). The patterns which are most often considered are the anti-ferromagnetic Spin Density Wave (SDW, opposite spins on different sublattices), and the Charge Density Wave (CDW, opposite charges on the two sublattices). At the level of the massless Dirac Hamiltonian with $N_f = 2$ fermion species all these patterns are energetically undistinguishable and are related by $U\lr{4}$ chiral rotations.

In order to make the fermionic operator $M$ invertible, one typically opens a small gap in the quasiparticle spectrum by breaking one of these symmetries, for example, by adding a perturbation of the form $m \Delta_{\psi} + m \Delta_{\chi}$ to the Hamiltonian (\ref{fermionic_hamiltonian1}). At the same time, such a perturbation induces a nonzero order parameter (SDW anti-ferromagnetism in this case). By extrapolating the results to $m = 0$ one can find the signatures of spontaneous symmetry breaking for the corresponding order parameter, as it is done in \cite{Buividovich:12:1,Buividovich:13:5,Smekal:13:1,Smekal:13:3}. An alternative and yet unexplored way to introduce the mass gap is to use lattice sizes ($L_x/L_y \neq 3/2$ \cite{Buividovich:12:1}) for which the Dirac points do not correspond to any discrete lattice momenta, so that no particular symmetry breaking pattern is favored.
 
First simulation results with the single-particle Hamiltonian (\ref{graphene_hamiltonian}) on sufficiently large hexagonal lattices (up to $24 \times 24$ elementary lattice cells) were reported in \cite{Buividovich:12:1}. Surprisingly, the measurements of both the chiral condensate (more precisely, the SDW order parameter) and the electric DC conductivity again indicated the value of the critical coupling $\alpha_c \approx 1.0 \pm 0.1$ significantly below the effective coupling constant $\alpha_{eff} = 2.2$ in clean suspended graphene.

\begin{figure*} 
  \centering
  \includegraphics[width=4.65cm]{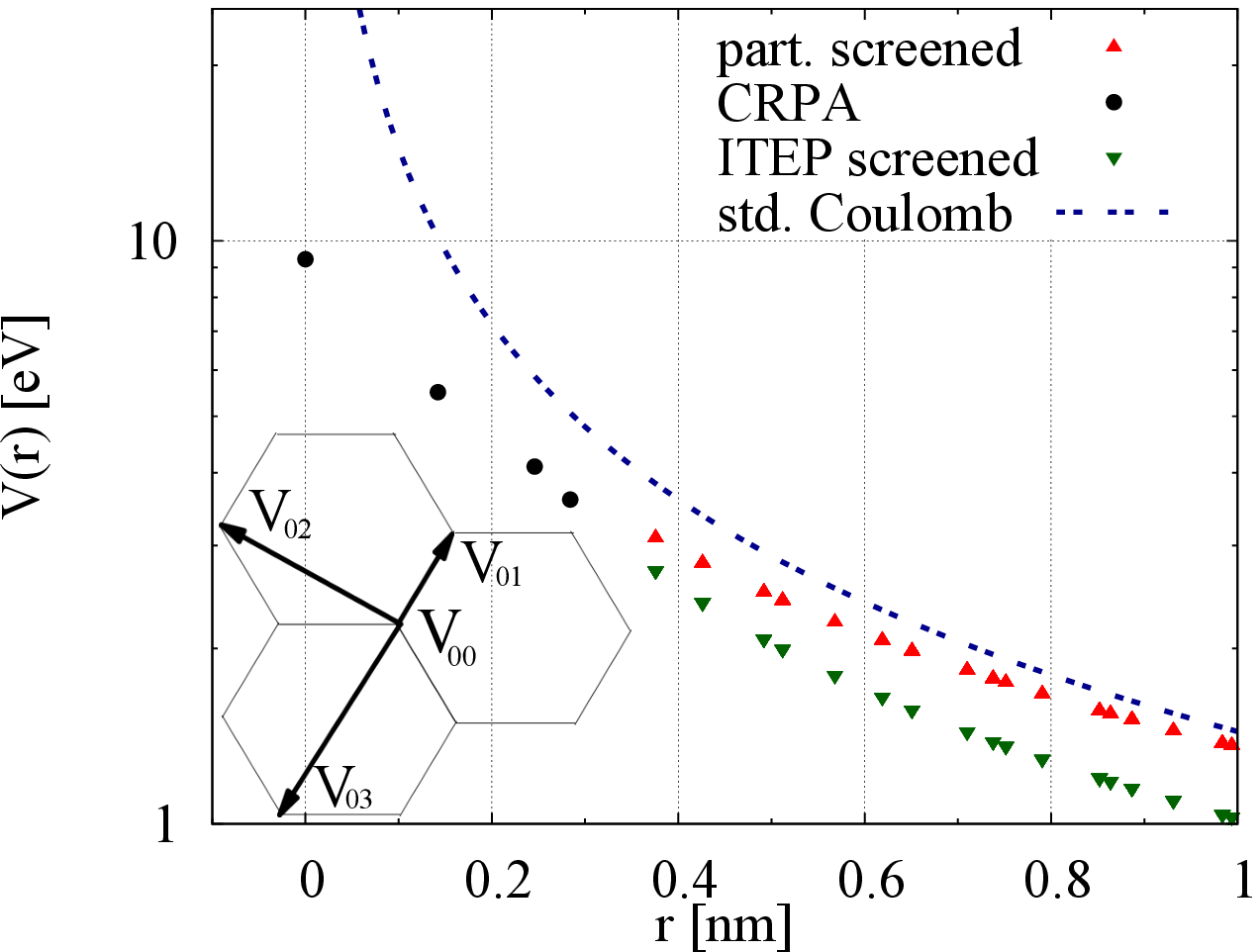}
  \includegraphics[width=5.0cm]{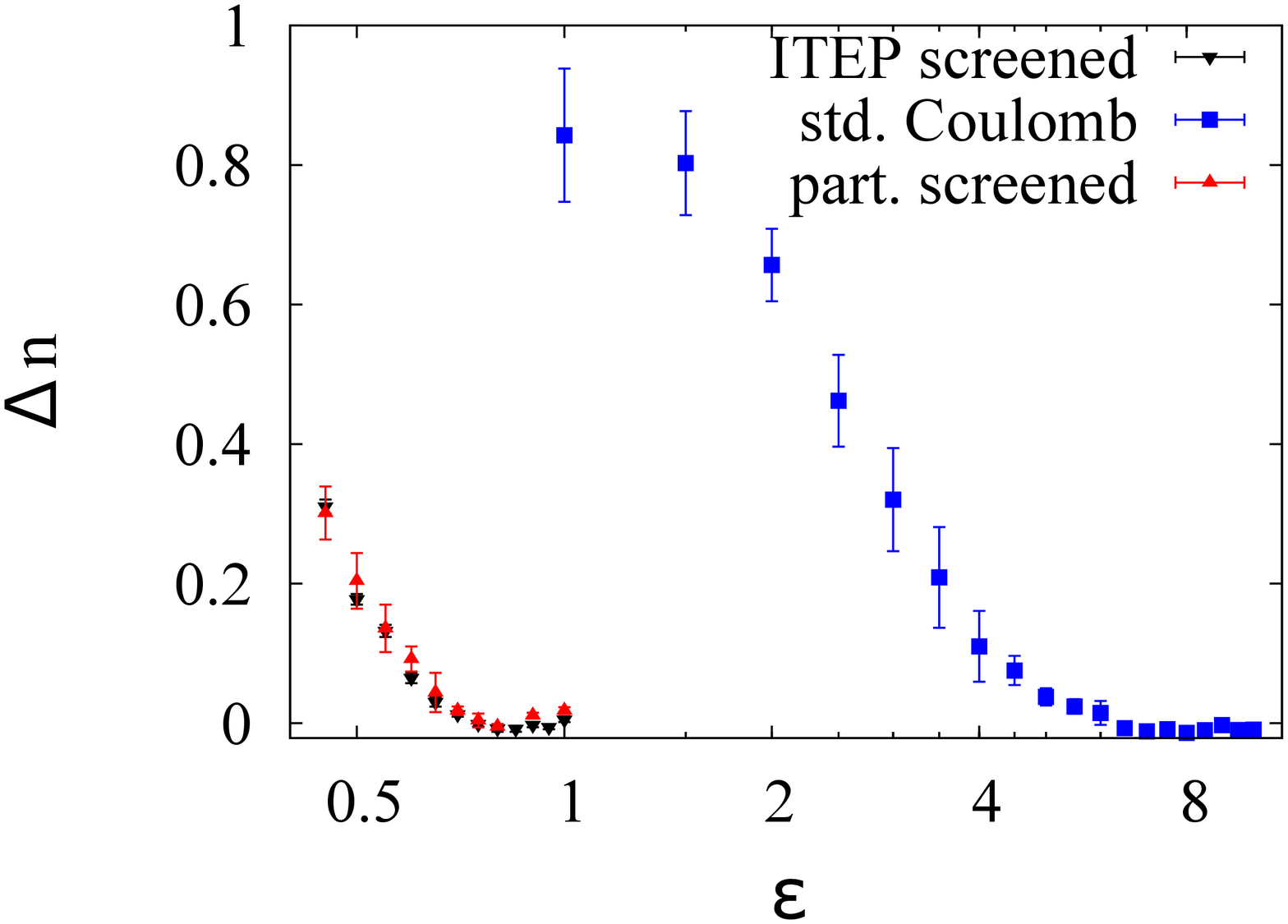}\\
  \caption{On the left: profiles of different inter-electron interaction potentials. On the right: anti-ferromagnetic order parameter as a function of interaction strength for the same set of potentials (dataset ``part. screened'' kindly provided by L.~{von Smekal} and D.~Smith \cite{Smekal:13:1,Smekal:13:3}).}
  \label{fig:potential_comparison}
\end{figure*}

 This long-standing puzzle on the insulating or semi-metal nature of the clean suspended graphene was only resolved when the effect of $\sigma$ orbitals on the inter-electron interactions was explicitly taken into account. As such, the conventional tight-binding model of graphene only describes the electrons on the $\pi$ orbitals of carbon. If one separately calculates how the electrons on $\sigma$ orbitals contribute to the screening of the Coulomb potential, Monte-Carlo simulations of the tight-binding model of $\pi$ orbitals do not result in any double counting. The contribution of $\sigma$ orbitals to the screening of the bare Coulomb potential within the distance of the few lattice spacings was calculated in \cite{Wehling:11:1} using the constrained Random Phase Approximation (cRPA). In \cite{Buividovich:13:5}, this potential was extrapolated to large distances by sewing it with the uniformly screened continuum Coulomb potential $V_{xy} = \alpha_{eff}/ (\epsilon_{\infty} |x - y|)$, $\epsilon_{\infty} = 1.36$. Such extrapolation is not quite correct, since due to the two-dimensional nature of graphene any electrostatic screening should be absent at large distances. Monte-Carlo simulations with the potential which smoothly interpolates between the screened short-distance potentials of \cite{Wehling:11:1} and the unscreened Coulomb potential $\alpha_{eff}/|x - y|$ were performed in \cite{Smekal:13:1,Smekal:13:3}, confirming practically the same critical value $\alpha_{c}$.

 On Fig.~\ref{fig:potential_comparison}, we compare the inter-electron interaction potentials used in \cite{Buividovich:12:1}, \cite{Buividovich:13:5} and \cite{Smekal:13:1,Smekal:13:3} with the continuum Coulomb potential. One can see that the contribution of $\sigma$ orbitals results in quite a significant screening of the Coulomb potential, especially at short distances. In particular, the on-site interaction potential is almost two times smaller. We note that since in \cite{Buividovich:13:5} the gauge fields were discretized on the hexagonal lattice, even the on-site interaction potential is finite, in contrast to the continuum Coulomb potential. On the other hand, from Fig.~\ref{fig:potential_comparison} one can see that the difference between the discretized and the continuum Coulomb potentials becomes quite small already at the distance of one bond length.

 In order to illustrate the influence of screening of the inter-electron potential on the spontaneous symmetry breaking in the SDW channel, we rescale all the interaction potentials by a factor $2/\lr{\epsilon + 1}$, where $\epsilon$ corresponds to a static dielectric permittivity of the fictitious substrate. On Fig.~\ref{fig:potential_comparison} on the right we compare the Monte-Carlo results for the anti-ferromagnetic SDW order parameter calculated with different inter-electron interaction potentials. We observe that for the screened potentials used in \cite{Buividovich:13:5} and \cite{Smekal:13:1,Smekal:13:3}, the onset of the spontaneous symmetry breaking is shifted to the region of unphysically small substrate dielectric permittivities $\epsilon < 1$ - for which the electrostatic interactions are stronger than in the vacuum! On the other hand, with the screened potential the suspended graphene with $\epsilon = 1$ is still within the semimetal phase, where the gap in the quasiparticle spectrum is absent and no symmetry is spontaneously broken.

 This finding is in perfect agreement with the experimental results of \cite{Elias:11:1,Elias:12:1}, and enables the quantitatively exact simulations of the suspended graphene. Let us stress that it is only possible to reach this agreement with experiment by using the tight-binding model of graphene on the hexagonal lattice, since the screened potentials of \cite{Wehling:11:1} only make sense on the hexagonal lattice. For staggered fermions on the square lattice, it might also be possible to mimic the effect of short-distance screening, but the simulation results anyway would not be quantitatively exact. This nicely illustrates the unique role of the tight-binding model (\ref{graphene_hamiltonian}) of graphene, for which the parameters are known very precisely, thus enabling very precise quantitative predictions from Monte-Carlo simulations - much like in the modern state-of-the-art lattice QCD simulations!

\section{Applications}
\label{sec:applications}

\subsection{Graphene in magnetic field}
\label{subsec:magnetic_field}

While the interaction strength in real graphene is not enough to trigger spontaneous symmetry breaking, the semimetal-insulator phase transition can still be shifted to the region of physical interaction strength by some external factors. One of the most obvious ones is the external magnetic field perpendicular to graphene plane. This scenario was extensively studied even before the actual experimental discovery of graphene \cite{Gusynin:94:1, Miransky:02:1}. In general it's quite similar to magnetic catalysis in QCD: chiral condensate is related to the density $\rho\lr{\lambda}$ of the eigenvalues of Dirac operator via Banks - Casher relation: $\langle \bar \psi \psi \rangle = \frac {\pi} {V} \rho\lr{\lambda \rightarrow 0}$. Since 2+1-dimensional fermions can not move along direction of perpendicular magnetic field, linear dispersion relation for massless Dirac fermions is reduced to series of  discrete relativistic Landau Levels $E_n = \pm v_F \sqrt{2 n e B /c}$, $n=0,1,2...$. Density of states contains now delta-function singularities for infinitely large sample, when we neglect boundaries. In the graphene tight-binding model relativistic Landau levels are still present well below van Hove singularities.  In particularly, zero Landau level causes delta-function in density of states at Fermi level.  If we look at relation between eigenstates of lattice fermionic operator (\ref{fermionic_matrix_op})  and eigenstates of one-particle Hamiltonian  $E_\xi$ \cite{Buividovich:12:1} in the absence of interaction:
\begin{equation}
\label{eigenstate}
\lambda (\omega, \xi) = 1- e^{i \omega \Delta \tau} (1 - E_\xi\Delta \tau), \quad  \omega=[0, \frac{2\pi}{\Delta \tau}],
\end{equation}
we immediately see that singularity in spectral density of one-particle Hamiltonian near $E_\xi=0$ leads to singularity of spectral density of the whole fermionic operator near $\lambda=0$. 

Shovkovy with coauthors \cite{Gusynin:94:1, Miransky:02:1} used Schwinger equation to study this system. They modelled graphene as 2 flavors of 2+1D Dirac fermions coupled with ordinary electromagnetic field. They showed appearance of chiral condensate even in infinitesimally small magnetic field $B$ at zero temperature.  Kharitonov \cite{Kharitonov:12:4} studied magnetic catalysis using renormalization group analysis and revealed rich phase diagram containing charge density wave (CDW), antiferromagnetic (AF) and ferromagnetic ordering and Kekule distortion. One can refer to the review \cite{Miransky:15:1} for the summary of all theoretical advances.

\begin{figure*} 
  \centering
  \includegraphics[width=5cm]{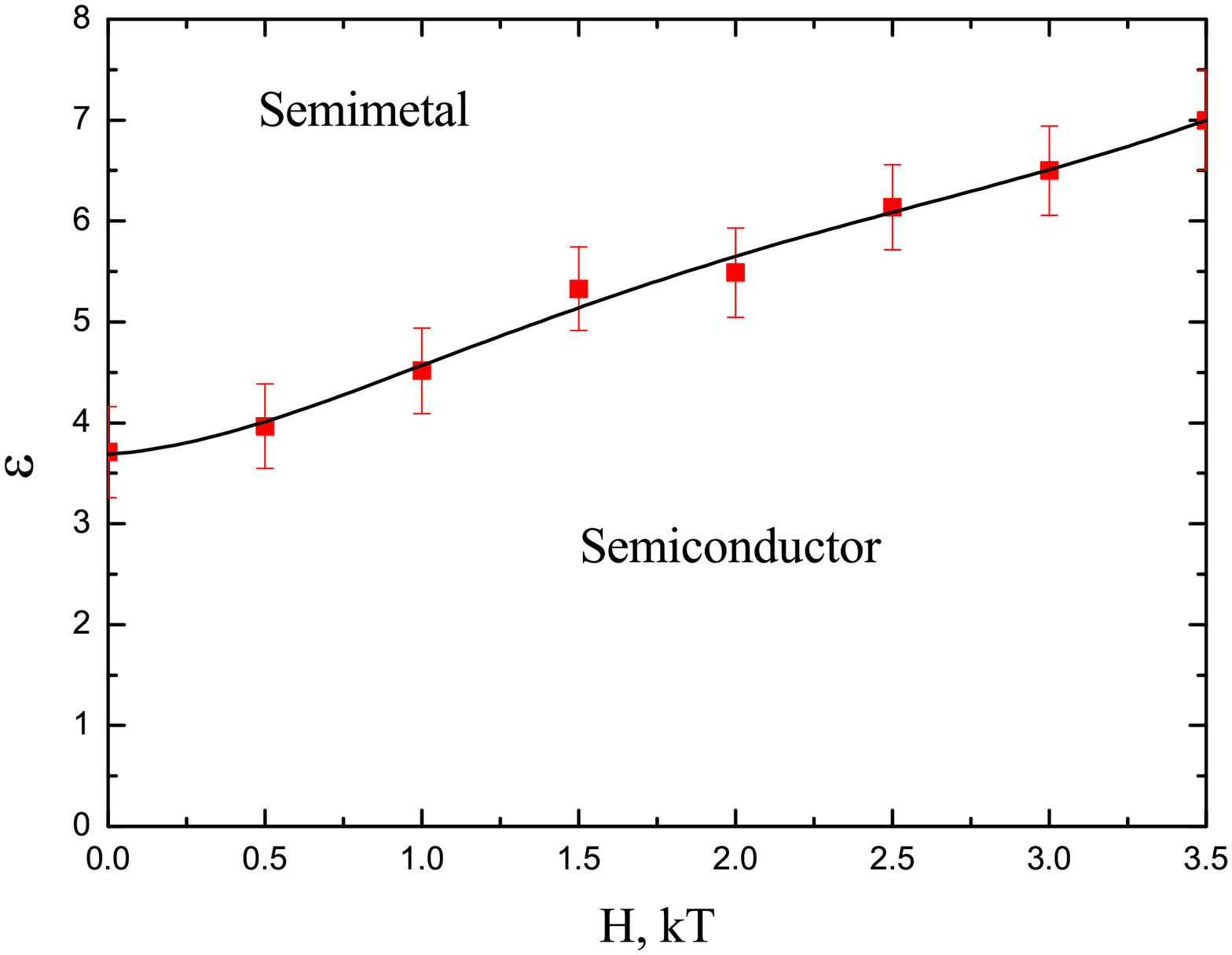}\\
  \caption{Phase diagram of 2+1D staggered fermions with instantaneous Coulomb interaction in external magnetic field. Semiconductor phase corresponds to the region with nonzero chiral condensate.}
  \label{fig:magnetic_phase_diagram}
\end{figure*}

Magnetic field can be easily introduced in lattice simulation with staggered fermions on the square lattice.  This setup was used in the papers \cite{ Cea:12:9, Ulybyshev:14:6, Winterowd:15:01}. In the paper \cite{Cea:12:9} compact $\lr{2+1}D$ electrodynamics was studied in external magnetic field. Thus not only fermions but also electromagnetic field was bounded to two-dimensional plane (thus electron-electron interaction was not of usual Coulomb form).  Linear dependence of chiral condensate on magnetic field was observed in the  weak coupling region, thus magnetic catalysis scenario was qualitatively approved.  In the paper \cite{Ulybyshev:14:6}  $\lr{2+1}D$ staggered fermions were coupled to usual $\lr{3+1}D$ electromagnetic field, in order to reproduce standard instantaneous Coulomb interaction.  Both weak coupling and strong coupling regions were studied. Linear dependence of chiral condensate on magnetic field did not appear in the weak coupling region, probably due to finite temperature effects. But magnetic catalysis was manifested in the shift of phase transition towards smaller interaction strength (see the phase diagram in Fig.~\ref{fig:magnetic_phase_diagram}).

In the recent paper \cite{Winterowd:15:01} the region of small coupling and strong magnetic field was studied. Again the appearance of chiral condensate was demonstrated in qualitative agreement with theoretical predictions. 

 Of course, modification of  electron-electron interaction at small distances was not taken into account in these papers and calculations can not be compared with experiment  directly. But since many theoretical predictions \cite{Gusynin:94:1, Miransky:02:1} were also based on pure Coulomb interaction, calculations with staggered fermions can be used as a benchmark for these results.  

\subsection{Magnetic moments in graphene with defects in crystal lattice}
\label{subsec:defects}
Another way to observe gap opening and phase transition in real graphene  is introduction of some kind of defects in crystal lattice. Hydrogen adatoms is one of the most advantageous ways for catalysis of phase transition.  They are also rather easy in simulations.
 Electronic structure in presence of such defects can be  explained by the fact that sp$^3$ state of carbon atom originated from its bond with a univalent adatom (like hydrogen) makes it unavailable for $p_z$ electrons ($\pi$-orbitals) of neighbouring carbons; for these electrons such atom is just cut from the lattice. Thus, hydrogen adatoms can be modelled by simple ``vacancies'' with hoppings to nearest neighbours are set to zero. It is well-known \cite{Peres:06:12}, that such ``vacancies'' generate sharp peak in the density of states near Fermi level.  Therefore all arguments from previous paragraph are valid again and we should expect sufficient enhancement of interaction effects. There were some studies of such systems using  Density Functional Theory \cite{Yazyev:07:1},  but they were limited to rather small sample size with periodical boundary condition. It  means that effectively adatoms were placed regularly on graphene sheet. HMC allowed to simulate large sample  with random (non-regular) distribution of defects \cite{Ulybyshev:15:6}.  Thus  energy levels can be estimated more realistically.  It is also possible to study mutual influence of two and more distant adatoms in various spatial configurations.
 
 \begin{figure*}
  \centering
  \includegraphics[width=5.0cm]{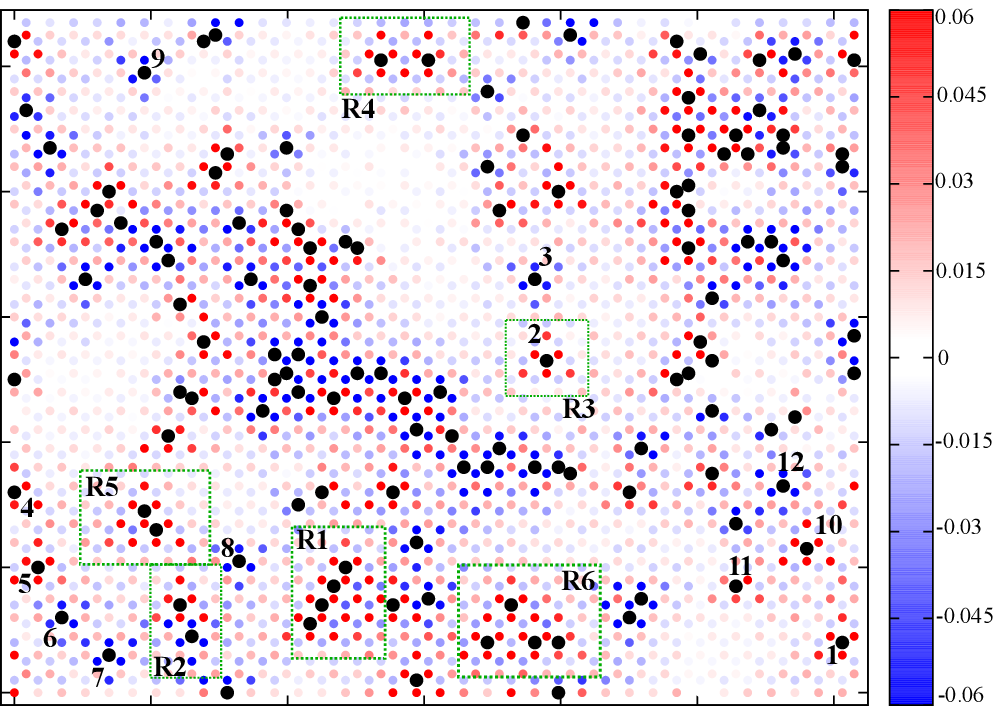}
  \includegraphics[width=4.9cm]{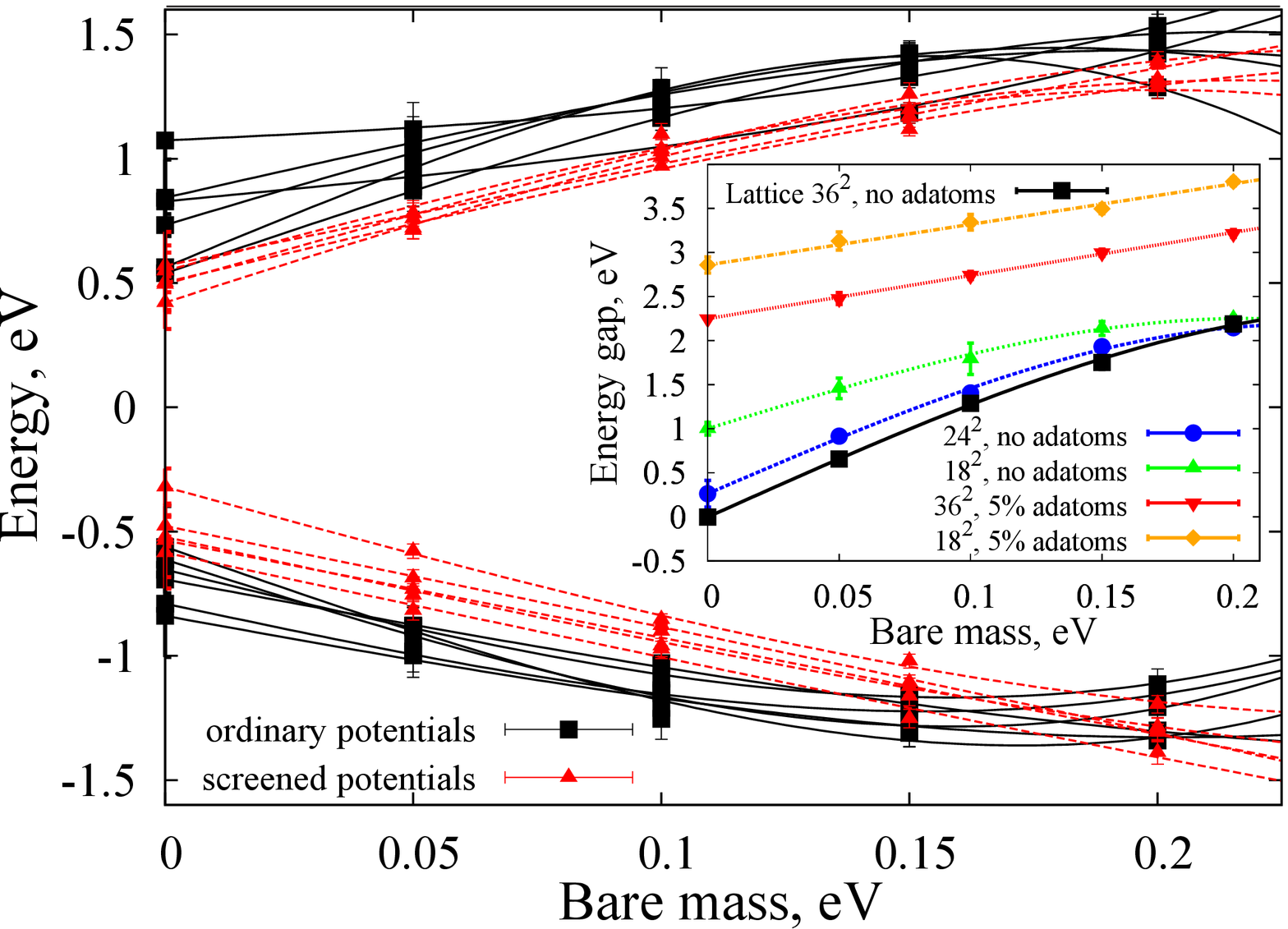}\\
    \caption{On the left: Distribution of average spin in presence of hydrogen adatoms (black points in the figure). Color scale corresponds to $\langle Sz \rangle$ at the given site. On the right: 1) Main plot: Energies of midgap states for two sets of inter-electron potentials. Ordinary potentials correspond to suspended graphene. Screened potentials correspond to graphene on boron nitride. Each state corresponds to one isolated vacancy marked with a number in the figure with spin distribution;  2) Inset: Energy gap between "normal" energy bands (former Dirac cones) measured for suspended graphene on different lattices in presence of adatoms and without them.  All values correspond to the K-point in Brillouine zone.  In all cases real physical situation is restored in the limit $m \rightarrow 0$.}
  \label{fig:adatoms_study}
\end{figure*}

 Distribution of electron spin is shown in the figure  \ref{fig:adatoms_study} on the left plot.  It is clearly seen that antiferromagnetic order is generated in the vicinity of adatoms. Moreover, one isolated adatom has nonzero average spin.  This spins tend to be parallel for adatoms at one sublattice and antiparallel for adatoms at different sublattices. If adatoms are placed equivalently at both sublattices, they generate opposite spin excess and thus the full spin will be close to zero. These measurements indeed show  catalysis of AF ordering.  It should be accompanied by appearance of mass gap. Corresponding calculations have also been performed in \cite{Ulybyshev:15:6}. Energy levels were extracted from exponential fit of the following correlator:
 
 \begin{equation}
 C (\tau) = \sum\limits_{x, y} \Tr \left({ \hat{a}^{\dag}_{x} \bar \psi(x) e^{-\tau \hat H} \hat{a}_{y} \psi(y) e^{- (\beta-\tau) \hat H} }\right).
\label{green_function}
\end{equation}
It is a  contraction of fermionic Green function with some guess for a wavefunction of corresponding energy state. 
The lowest energy band is formed by midgap states which are located inside the large gap opened between "normal" energy bands (former Dirac cones). Wavefunctions of midgap states are  concentrated near relatively isolated adatoms.  Examples for such adatoms are  marked with black numbers in the figure  \ref{fig:adatoms_study} on the plot with spin distribution.  Resulting energies are shown in figure \ref{fig:adatoms_study} on the right plot. 

Final physical gap can be obtained in the zero bare mass limit.  For 5\% concentration of hydrogen adatoms it is around 1.1 eV for suspended graphene and around 0.74 eV for graphene on boron nitride substrate which screens Coulomb interaction at large distances.
Measurements of magnetic moments of various configurations of adatoms demonstrated large influence of distant defects on each other.  Further study of this influence seems to be important task, especially taking into account possible effects of clusterization.  One needs to simulate large samples to study this effects and HMC gives us good possibility to perform such calculations.

\subsection{Nonperturbative renormalization from lattice calculations}
\label{subsec:renornalization}

It is reasonable to assume that because of strong interaction, the obvervables of the graphene theory are strongly renormalized as compared to non-interacting theory. For example, the leading perturbative corrections to the Fermi velocity \cite{Gonzalez:94:3} lead to its logarithmic renormalization. One can expect that higher order renormalization leads to further considerable change of the bare value of the Fermi velocity but existing experimental measurements \cite{Elias:11:1} are in good agreement with the first-order perturbation theory improved by the one-loop expression for the dielectric permittivity of graphene.  Recently it was shown that the next-to-leading order corrections in the random phase approximation (RPA), are small relative to the leading-order RPA results \cite{Sodemann:12:11}. Nevertheless it is not clear what happens with perturbative corrections after the next-to-leading order. 

Renormalization of $v_F$ is also interesting due to another reason: coincidence between calculated value of Fermi velocity and the one measured in experiment can serve as a criterion for tuning the effective coupling constant in simulations with staggered fermions. After this phenomenological tuning the simulations with staggered fermions are better matched with real physics. This program was carried out in the paper \cite{Drut:13:4}, renormalized $v_F$ was obtained  by fitting Monte-Carlo data with some guess for fermionic propagator.

Closely related to calculation of $v_F$ renormalization is the study of dispersion relation in presence of strong electron-electron interaction. It can be done using HMC calculation of correlator (\ref{green_function}) using complex exponent $\exp{(i\vec k \vec x)}$ as a guess for the wavefunction. In case of suspended graphene with or without adatoms it was done in \cite{Ulybyshev:15:6} for K-point in Brillouine zone (see inset in the right plot on the Fig. \ref{fig:adatoms_study}). Thus the gap between main energy bands can be estimated. The same study was performed in \cite{Drut:13:4} for the whole possible momenta in carbon nanotubes. Interesting and still unclear question is connection between energy gap in nanotubes \cite{Luu:15:11} and the gap appeared in the K-point of clean suspended graphene in the case of measurements on relatively small lattices, see Fig. \ref{fig:adatoms_study}. Probably, finite-size effects in suspended graphene are quite significant (thus we need simulations of larger samples) and are not well understood.

Another interesting observable for which one can expect sizable renormalization is the optical conductivity $\sigma(\omega)$. 
The $\sigma(\omega)$ descibes electric current in graphene resulting from external electric field $j(\omega)=\sigma(\omega) E(\omega)$. 
In the non-iteracting theory of Dirac quasiparticles the conductivity does not depend on frequency in the limit of zero temperature. It equals to $\sigma(\omega) = \sigma_0 = \frac {e^2} {4 \hbar}$.
The departure of the $\sigma(\omega)$ from the value $\sigma_0$ at zero temperature can be attributed to 
the interaction between quasiparticles. Experimental measurement of the optical 
conductivity didn't find \cite{Nair:08} deviation from the result of non-iteracting theory. 

There are a lot of papers devoted to the leading-order perturbative correction to the value of  $\sigma(\omega)$ which present different results (for short review see \cite{Boyda:16:02} and references therein). Discussion is concentrated mainly around the value of constant $C$ in the general formula for  optical conductivity renormalization:
\begin{equation}
\frac {\sigma(\omega)} {\sigma_0} = 1+ C \alpha_{eff} +O (\alpha_{eff}^2),
\label{sigma_renorm}
\end{equation}
where $\alpha_{eff}$ is logarithmically dependent on frequency.
There is still large disagreement in theoretical predictions: while some of papers \cite{Herbut:08:1} predict large renormalization $C=0.2...0.5$, another papers \cite{Mishchenko:08:1} claim that $\sigma (\omega)$ is almost stable with $C \approx 0.01$.

Recently optical conductivity of graphene was studied using HMC on hexagonal lattice \cite{Boyda:16:02}. $\sigma(\omega)$ was extracted from Euclidean current-current correlator. Green-Kubo relation was solved using  Backus-Gilbert method, recently adopted for similar tasks in lattice quantum chromodynamics \cite{Meyer:15:9}.  This method is based on the notion of resolution functions. As a result we obtain convolution of optical conductivity with resolution functions $\delta(\omega_0, \omega)$:
$\bar \sigma(\omega_0) = \int_0^\infty  \delta(\omega_0, \omega) \sigma(\omega) d\omega$.
It makes this method especially suitable for definition of the value of spectral function at some plateau (exactly the case of optical conductivity which forms  plateau at small frequencies). If parameters of simulation (temperature, etc. ) are tuned to make the width of resolution functions  smaller than the width of plateau, results become very stable and independent on any details of the algorithm.  This is advantageous in comparison with Maximal Entropy method (MEM). On the other hand MEM works better in definition the positions and widths of resonances. Backus-Hilbert method can even miss the resonance if resolution function is not enough narrow.

Results of  Monte-Carlo calculations doesn't show any renormalization even for suspended graphene thus strongly supporting theoretical claims of very small constant $C$ in (\ref{sigma_renorm}).

\subsection{Bilayer graphene}
\label{subsec:bilayer}

Bilayer graphene was studied in two ways using HMC. First of all, voltage-biased bilayer graphene was studied in \cite{Hands:13:6} using simple model of 4 flavors of 2+1D Dirac fermions. They were simulated by two copies of ordinary staggered fermions coupled via instantaneous Coulomb interaction. Hopping between layers was not taken into account.  Voltage bias between layers was introduced through the opposite chemical potential for two flavours of staggered fermions.  In this setup both Dirac cones are shifted in opposite directions thus we have non-vanishing  density of states at Fermi level. According to the reasons discussed in paragraph \ref{subsec:magnetic_field} a new condensate can appear. It is  composed from particles in one layer and holes in the other.
Indeed, this type of condensation was observed while ordinary chiral condensate was strongly suppressed. 

Other attempt to study bilayer graphene was presented in \cite{Nikolaev:14:12}. Here the microscopical tight-binding model for AA-stacked bilayer graphene was studied using the same technique that was already used for monolayer graphene. Antiferromagnetic condensation was observed, in qualitative agreement with previous theoretical predictions \cite{Rozhkov:12}. Nevertheless, some deviations were also found. They  probably relate to the role of long-range Coulomb tail which works against formation of AF order. 

\section{Future perspectives: topological insulators and Dirac and Weyl semimetals}
\label{sec:outlook}

\subsection{Topological insulators}
\label{subsec:TIs}

 Topological insulators are materials which have gapped energy spectra in the bulk and are in this respect similar to the conventional insulators, but can host massless excitations on the boundary of a sample which cause nonzero boundary conductivity. At low energies these boundary excitations can be described as Weyl fermions with an effective speed of light being equal to the Fermi velocity $v_F \ll c$. The existence of Weyl quasi-particles at the boundary is topologically protected against moderate perturbations (such as disorder or inter-electron interactions) by the time-reversal symmetry \cite{Kane:05:1, Kane:05:2, Fu:07:1, Fu:07:2}. Topological insulator phases can exist both for two-dimensional \cite{Kane:05:1, Kane:05:2} and three-dimensional materials \cite{Fu:07:1, Fu:07:2} with strong spin-orbital coupling (SOC).

 One of the simplest models of 2D topological insulators is the Kane-Mele model \cite{Kane:05:1, Kane:05:2} with the following single-particle Hamiltonian:
\begin{eqnarray}
\label{KM_hamiltonian}
  h^{\psi, \chi}
  =
  \sum\limits_{<x,y>} \kappa \delta_{x,y}
 +
 \sigma \, \sum\limits_{<<x,y>>} \pm i \kappa' \, \delta_{x, y}
\end{eqnarray}
where the summation indices $<x, y>$ and $<<x,y>>$ denotes summation over all pairs of nearest-neighbor and next-to-nearest neighbor sites of the hexagonal lattice, respectively, and the sign before the next-to-nearest-neighbor hopping amplitude $i \kappa'$ depends on the hopping direction in a non-trivial way \cite{Kane:05:2}. The first summand in (\ref{KM_hamiltonian}) is just the tight-binding model of graphene. The second summand explicitly depends on the electron spin $\sigma = \pm 1/2$ and thus describes the spin-orbital coupling. Remembering that in graphene particles and holes can be associated with spin up and spin down states, we assume $\sigma = +1/2$ in $h^{\psi}$ and $\sigma = -1/2$ in $h^{\chi}$. This SOC generates a topological energy gap, for which the signs of the masses are different for different Dirac points and spin orientations. While the Kane-Mele model (\ref{KM_hamiltonian}) is not directly applicable to real graphene, where the SOC is too small, it can be used to make qualitative predictions on the effects of disorder and other bulk and edge instabilities in real materials such as HgTe/CdTe quantum wells \cite{Konig:07:1}.

One possible bulk instability is the competition between the topological mass gap induced by SOC and the Dirac mass term which results from spontaneously generated anti-ferromagnetic order at sufficiently strong on-site interactions\cite{Meng:10:1, Hohenadler:12:1, Zheng:11:1}. Recent HMC simulations of graphene (see Section \ref{sec:graphene_history}) and the mean-field arguments of \cite{Raghu:08:1} suggest that non-local interactions can strongly affect the phase diagram of the Kane-Mele model - a conjecture which seems natural to check using Monte-Carlo simulations.

 In three dimensions, probably the simplest theoretical model of topological insulators is the single-particle Wilson-Dirac Hamiltonian on 3D cubic lattice \cite{Fu:07:1,Zhang:09:1,Araki:13:2}, which reads in the momentum space representation:
\begin{eqnarray}
\label{WD_hamiltonian}
 h_{WD}\lr{k} 
 =
 \kappa \sum\limits_{i=1}^{3} \alpha_i \sin\lr{k_i} + \kappa \beta \lr{m_0 + 2 r \sum\limits_{i=1}^{3} \sin^2\lr{k_i/2}} ,
\end{eqnarray}
where $\alpha_i$ and $\beta$ are the Dirac $\alpha$- and $\beta$- matrices, $\kappa$ is the dimensionful hopping parameter, $r$ is the Wilson parameter which roughly corresponds to SOC in real 3D topological insulators and $m0$ (with $-2 r < m_0 < 0$) is the topological mass term \cite{Fu:07:1, Zhang:09:1, Araki:13:2}. The Hamiltonian (\ref{WD_hamiltonian}) can be obtained as a low-energy approximation to the tight-binding models of the real-world topological insulators such as $Bi_2 Se_3$, $Bi_2 Te_3$ and $Sb_2 Te_3$ \cite{Zhang:09:1}. 

Inter-electron interaction effects in 3D topological insulators and the SOC can be equally important in compounds involving $5d$ transition metals, and their interplay might result in interesting new topologically insulating phases. For example, the surface of a topological insulator with sufficiently strong on-site interactions might support gapless spinon excitations while being electrically insulating \cite{Pesin:10:1}. In combination with strong on-site interactions, nearest-neighbor interactions might also induce an effective SOC even if it is absent in the non-interacting Hamiltonian \cite{Zhang:09:2, Kurita:11:1}, and thus lead to the interaction-induced topological insulator phase.

Fortunately, the time-reversal symmetry of the time-reversal-invariant topological insulators is not broken by purely electrostatic interactions in the absence of external magnetic fields. This is enough to ensure the absence of the sign problem in determinantal Monte-Carlo simulations, which opens a way to the first-principle studies of the novel interaction-induced topological phases. 

For the Kane-Mele model (\ref{KM_hamiltonian}), the matrix elements of the single-particle Hamiltonians $h^{\psi}$ and $h^{\chi}$ (and, correspondingly, of the fermionic operators $M^{\psi}$ and $M^{\chi}$) are complex conjugate. This implies the positivity of the determinant in (\ref{partition_function_HS}), and, in addition, allows to avoid explicit determinant calculation by using the $\Phi$ algorithm (see Section \ref{sec:hmc_intro}). The effect of long-range Coulomb interactions on the phase diagram of the Kane-Mele model has been studied recently using the determinantal Monte-Carlo algorithm with the exact calculation of the determinants, which limited lattice sizes to $18 \times 18$ unit cells \cite{Herbut:14:1}. It was found that, similarly to the case of graphene, the long-range potential shifts the critical coupling of the transition from the topological quantum spin Hall insulating state to the non-topological anti-ferromagnetic insulating state to significantly higher values.

For the Wilson-Dirac Hamiltonian (\ref{WD_hamiltonian}) the time-reversal symmetry ensures that the eigenvalues of the corresponding fermionic operator appear either in complex conjugate pairs or in pairs of doubly degenerate real values (Kramers degeneracy), which implies the positivity of the determinant even for $N_f = 1$ fermion flavor. This degeneracy has no analogue in lattice QCD, where space-like gauge fields always break time-reversal invariance and hence introduce the (real) sign problem into simulations with $N_f = 1$ fermion flavour.

\subsection{Dirac and Weyl semimetals}
\label{subsec:Dirac_Weyl}

 Weyl semimetals are characterized by several Fermi points in the Brillouin zone, near which electronic excitations can be described as either left- or right-handed Weyl fermions in $\lr{3 + 1}$ dimensions, again moving with a Fermi velocity $v_F \ll c$. Separated Weyl points in Weyl semimetals are topologically protected against moderate perturbations by a nontrivial Berry flux in momentum space \cite{Wan:11:1,Kim:12:1}, in contrast to Dirac semi-metals, for which Weyl points of opposite chiralities coincide and even an infinitely small perturbation (e.g. a Dirac mass term) can open the gap. The vanishing of the total flux in the compact Brillouin zone \cite{Nielsen:81:1} implies an equal number of left- and right-handed Weyl nodes.

 A simple but quite realistic lattice model of Weyl semimetals can be obtained from the model Hamiltonian (\ref{WD_hamiltonian}) by tuning the Dirac mass $m_0$ to the critical value (e.g. $m_0 = 0$) separating the topological insulator and the trivial insulator phases and introducing time-reversal and/or parity-breaking perturbations of the form \cite{Sekine:13:1, Vazifeh:13:1, Hosur:13:1}
\begin{eqnarray}
\label{WD_hamiltonian_WSM}
 {\delta h}\lr{k} = b_i \Sigma_i + \mu_A \gamma_5
\end{eqnarray}
where $\Sigma_i = -i\epsilon_{ijk} \alpha_j \, \alpha_k/2$ is the spin operator and $\gamma_5 = -\beta \alpha_1 \alpha_2 \alpha_3$ is the generator of chiral rotations. These perturbations shift the positions of otherwise coinciding left- and right-handed Weyl nodes to different momenta $\vec{k}^{\pm} \sim \pm \vec{b}$ and energies $E^{\pm} \sim \pm \mu_A$. Nonzero $b_i$ breaks time-reversal invariance (but not parity) and can be interpreted as a magnetic doping of 3D topological insulator \cite{Sekine:13:1}. The chiral chemical potential $\mu_A$ describes material with chirality imbalance (different numbers of left- and right-handed fermions), which can be generated by ``chirality pumping'' in parallel electric and magnetic fields. 

Mean-field analysis of \cite{Wang:13:1, Wei:12:1, Sekine:13:1}, strong-coupling expansion \cite{Sekine:13:1} as well as the systematic renormalization-group analysis \cite{Maciejko:13:1} have revealed an interesting possibility of spontaneous chiral symmetry breaking in Weyl semimetals with $b_i \neq 0$ and sufficiently strong on-site interactions, accompanied by the appearance of massless ``axion'' Goldstone modes (or simply pions from the point of view of lattice QCD) which are the fluctuations of the phase of the chiral condensate. Here, however, more efforts are required to obtain the first-principle results numerically, since at $b_i \neq 0$ the determinantal Monte-Carlo has a sign problem due to the explicit breaking of time-reversal invariance. However, at $b = 0$ nonzero chiral chemical potential $\mu_A$ does not break the time-reversal invariance, and hence simulations at nonzero $\mu_A$ are possible even with $N_f = 1$ flavor of Wilson-Dirac fermions with purely electrostatic interactions (see \cite{Yamamoto:11:2,Braguta:15:1} for recent lattice QCD simulations with chiral chemical potential).

\subsection{Lattice strong-coupling expansions for condensed matter systems}
\label{subsec:sc_expansion}

 An interesting alternative approach to the study of strong electron correlations in graphene, topological insulators and Dirac/Weyl semi-metals \cite{Araki:10:1,Araki:10:2,Araki:13:1,Araki:13:2,Araki:13:3,Sekine:13:1} is based on the lattice strong-coupling expansion \cite{CreutzLGT,PolyakovGaugeStrings}. The basic idea is to consider the Hubbard-Stratonovich field $\phi_x\lr{\tau}$ as a compact time-like gauge field (indeed, it couples to fermions as a compact gauge field), and to replace the quadratic action $S_{HS}\lrs{\phi_x} = 1/2 \sum\limits_{x,y} \phi_x V^{-1}_{x y} \phi_y$ with the action of the compact QED $S_{cQED} = -V^{-1} \sum\limits_{x,\mu} \cos\lr{\phi_x - \phi_{x + \hat{\mu}}}$. With the compact gauge action, one can perform the conventional lattice strong-coupling expansion in powers of $V^{-1}$.

In \cite{Araki:10:1}, spontaneous chiral symmetry breaking in graphene was demonstrated using the strong-coupling expansion. In \cite{Araki:13:1}, this method was applied to the Kane-Mele model (\ref{KM_hamiltonian}), and a novel tilted anti-ferromagnetic phase was predicted. In \cite{Araki:13:2}, strong-coupling expansion was applied to the three-dimensional topological insulator as described by the Wilson-Dirac Hamiltonian (\ref{WD_hamiltonian}), and the existence of topologically nontrivial state in the infinite coupling limit was demonstrated. In \cite{Sekine:13:1}, Weyl semimetals with $b \neq 0$ and $\mu_A = 0$ were studied using this technique, and a novel Weyl semimetal phase with spontaneously broken parity was predicted.

 Due to the condensation of monopoles which is a specific feature of compact QED \cite{Polyakov:75:1}, in the strong-coupling regime with large $V$ the compact QED action leads to confining linearly rising potential between static charges (in the absence of dynamical fermions), which is quite different from realistic inter-electron interactions in crystalline solids. Thus while the strong-coupling expansion is an interesting alternative to mean-field calculations, which in particular is far less affected by sign problems, strictly speaking it is not in one-to-one relation with the original many-body Hamiltonian (\ref{fermionic_hamiltonian1}), and probably an additional theoretical work is required to turn it into a tool for first-principle calculations.

\section*{Acknowledgments}

 The authors acknowledge fruitful and stimulating discussions with F.~Assaad, M.~I.~Katsnelson, L.~{von Smekal} and D.~Smith. P.B. is supported by the S.~Kowalevskaja award from the Alexander von Humboldt foundation. The work of M.U. is supported by the grant BU 2626/2-1 from the Deutsche Forschungsgemeinschaft (DFG).

\end{document}